\documentclass[12pt, a4paper]{article}
\usepackage{amsmath,amsfonts}
\usepackage{hhline}
\usepackage{indentfirst}
\usepackage{caption3}
\usepackage[pdfborder={0 0 0}]{hyperref}
\usepackage{tikz}
\usetikzlibrary{positioning}
\makeatletter
\renewcommand{\@oddhead}{Romanov V. F. (\href{mailto:romvf@mail.ru}{romvf@mail.ru})\hfil \thepage}
\makeatother

\begin{document}
\thispagestyle{empty}
\begin{center}
{\bf Non-Orthodox Combinatorial Models Based on Discordant Structures}
\end{center}
\begin{center}
{\bf Romanov V. F. (\href{mailto:romvf@mail.ru}{romvf@mail.ru})}
\end{center}
\begin{center}
Vladimir State University
\end{center}

\begin{abstract}
This paper introduces a novel method for compact
representation of sets of
$n$-dimensional binary sequences in a form of compact triplets  structures (CTS), supposing both logic and arithmetic interpretations of data. Suitable illustration of CTS application is the unique graph-combinatorial model for the classic intractable 3-Satisfiability problem and a polynomial algorithm for the model synthesis. The method used for Boolean formulas analysis and classification by means of the model is defined as {\it a bijective mapping principle for sets of components of discordant structures to a basic set.} The statistic computer-aided experiment showed efficiency of the algorithm in a large scale of problem dimension parameters, including those that make enumeration procedures of no use.

The formulated principle expands resources of constructive approach to investigation of intractable problems.

\end{abstract}

\textbf{1. Introduction. Tabular formulas}

A large number of discrete optimization problems are combinatorial and certain are intractable. Analysis and classification of these problems often involve reducibility methods based on models using special constructive components. By this reason, new research results on models, properties, computational techniques, and algorithms for some selected intractable problems often assume generalization.

In this paper a non-orthodox graph-combinatorial model for the classic 3-Satisfiability problem is presented. Efficient implementation of the algorithm based on the model leads to a polynomial running time for Boolean formula classification in a wide range of the problem dimension parameters.

The 3-Satisfiability (hereinafter 3-{\it SAT}) problem statement: for given  $m$  elementary disjunctions $C_1, C_2, \,\dots,\, C_m$, each containing exactly 3 literals, referring to Boolean variables $x_1, x_2, \,\dots,\, x_n$, determine, whether the formula $$F=C_1\wedge C_2\wedge \cdots \wedge C_m$$  is satisfiable or not.

We will use for the formula presented in the conjunctive normal form (CNF) a specific recording mode---the form of a table ({\it a tabular formula}), containing $n$ columns, noted by names of the variables, and  $m$ lines, each presenting the term $C_i$ by 0-1 sequence: 0  written in column  $j$  and line  $i$  marks the occurrence of $x_j$   in the term $C_i$ without negation, 1---with the sign of negation.
So, the tabular representation of the formula
$$
F=(\overline{x}_1 \vee x_2 \vee \overline{x}_4)\wedge
(x_2 \vee x_3 \vee \overline{x}_5)\wedge
(\overline{x}_3 \vee \overline{x}_4 \vee x_5)
$$ is as follows:

$$
\begin{array}{ccccccc}
&&F\\
x_1& x_2& x_3& x_4& x_5\\[3pt]
1&0&&1&\\
&0&0&&1&\\
&&1&1&0&.
\end{array}
$$

It is obvious that if 0 and 1 denote truth values: {\it false} and {\it true}, respectively (regardless of the denotation taken for the tabular formula\footnote{Actually we use dual interpretation of values 0 and 1 depending on a context.}), $F = 1$ {\bf for those and only for those sets of truth values, which do not contain any line from the tabular formula as a subset.}

\medskip
\textbf{2. Structures of compact triplets}

We consider for beginning a tabular formula consisting of terms in which three literals form {\it compact triplets (CT)}, that is  $\langle l_j$,~ $l_{j+1}$,~ $l_{j+2}\rangle$ sequences, where $l_j \in \{ x_j, \ \overline{x}_j\}$, $1\le j\le (n-2)$. We name such a formula a {\it CT formula {\rm (or} CTF}). The idea of CTF resolving is to transform the CTF to a structure of compact triplets (abbreviations: a {\it CT structure {\rm or} CTS}). The elements of the tabular CTS are lines, viz.\ {\it compact triplets} of variables' values located at  $n-2$  {\it tiers.} The tiers include variables numbered as 1,~2,~3; 2,~3,~4; \dots ; $n-2,~n-1,~n.$  Any CTS is composed of the triplets that are absent in the corresponding CTF, at each tier respectively. Generally, each tier contains a maximum of 8 binary lines. The final step of the CTS construction is a {\it clearing} procedure: removal from the tiers of {\it non-compatible lines}, i.e. the lines which cannot be adjoined to at least one line of each adjacent tier on condition that two values of variables written in succession coincide. The remaining lines are {\it compatible} and form sequences of length $n$  by means of {\it adjoining} operation (based on coincidence described above) applied to the pairs of lines from the tiers 1---2, 2---3, $\cdots,$ ($n-3$)---($n-2$). It is obvious that the CTF---CTS transformation is polynomial in terms of the algorithm complexity.

If at least one tier of the CTS turns out to be empty, the whole structure is declared an empty set of lines (or {\it an empty structure}) and the formula $F$ is declared a contradiction.
The CTS containing $n-2$ tiers can be formed if and only if $F$ is satisfiable. In fact, of a total of $2^n$ sequences of length $n$ all those and only those have been removed that include, as a subsequence, at least one line of the table representing formula $F$. Hence the CTS contains as the sequences of length $n$ all sets of truth values at which  $F$  is true (called {\it satisfying sets}).  Thus, {\it the very fact of existence of the CTS including  $n-2$  tiers means  that  $F$  is a satisfiable formula.}

Let us say that a structure of compact triplets is complete if each its tier contains 8 possible combinations of the binary values; such a structure represents the totality of $2^n$ satisfying sets ({\it SS}).

For example, the transformation of CTF  $F_1$  leads to the CTS  $Z$; the intermediate structure $Z^*$  still contains non-compatible lines (marked ``{\bf --}").

$$
\begin{array}{ccccccccccccccccccccccc}
&& F_1 &&&&& && Z^* &&&&& &&& Z &&\\[3pt]
x_1&x_2&x_3&x_4&x_5&&&x_1&x_2&x_3&x_4&x_5&&&&x_1&x_2&x_3&x_4&x_5\\[3pt]
0&0&0&&&&& 0&1&0&&&-&&& 0&1&1&&\\
0&0&1&&&&& 0&1&1&&&&&& 1&0&0&&\\
1&0&1&&&&& 1&0&0&&&&&& &1&1&0&\\
1&1&1&&&&& 1&1&0&&&-&&& &0&0&1&\\
&0&0&0&&&& &0&0&1&&&&& &&1&0&1\\
&1&0&0&&&& &0&1&0&&-&&& &&0&1&1\\
&1&0&1&&&& &0&1&1&&-&&& &&&&\\
&1&1&1&&&& &1&1&0&&&&& &&&&\\
&&0&1&0&&& &&0&0&0&-&&& &&&&\\
&&1&0&0&&& &&0&0&1&-&&& &&&&\\
&&1&1&1&&& &&0&1&1&&&& &&&&\\
&&&&&&& &&1&0&1&&&& &&&&\\
&&&&&&& &&1&1&0&-&&& &&&&\\
\end{array}
$$

Thus, analyzing $Z$, we fix two satisfying sets for $F$: 01101, 10011.

\medskip
{\bf Operations on CTS set}. We define for a system of CT structures, based on the {\it fixed} numbering of $n$ variables, a notion of {\it equivalence} and three operations: {\it union, intersection {\rm and} concretization}. The equivalence of two CT structures $(S_1=S_2)$ denotes equivalence of the sets of lines at all tiers with the identical numbers. Union and intersection of CTS's are similar to the same-name operations of the set theory, provided that the elements of sets (as applied to CTS-operands) are the lines of the tiers with the identical numbers; the clearing procedure of the resulting CTS terminates the intersection. The operation examples are presented below.

$$
\begin{array}{cccccccccccccccccccc}
&& S_1 && &&&&&&&& && S_2 &&\\[5pt]
x_1&x_2&x_3&x_4&x_5&&&&&&&&x_1&x_2&x_3&x_4&x_5\\[3pt]
0&1&0&&&&&&&&&& 0&1&1&&\\
0&1&1&&&&&&&&&& 1&0&0&&\\
&1&0&1&&&&&&&&& &1&1&0&\\
&1&1&0&&&&&&&&& &0&0&1&\\
&&0&1&1&&&&&&&& &&1&0&1\\
&&1&0&0&&&&&&&& &&0&1&1\\
&&1&0&1&&&&&&&& &&&&\\
\end{array}
$$
\par\medskip
$$
\begin{array}{ccccccccccccccccccccc}
S_3&= &S_1&\cup &S_2&&&&&&&& S_4&=&S_1&\cap &S_2\\[3pt]
x_1&x_2&x_3&x_4&x_5&&&&&&&&x_1&x_2&x_3&x_4&x_5\\[3pt]
0&1&0&&&&&&&&&& 0&1&1&&\\
0&1&1&&&&&&&&&& &1&1&0&\\
1&0&0&&&&&&&&&& &&1&0&1\\
&1&0&1&&&&&&&&& &&0&1&1&-\\
&1&1&0&&&&&&&&& &&&&\\
&0&0&1&&&&&&&&& &&&&\\
&&0&1&1&&&&&&&& &&&&\\
&&1&0&0&&&&&&&& &&&&\\
&&1&0&1&&&&&&&& &&&&\\
\end{array}
$$

In $S_4$ the last line is deleted according to the clearing procedure.

Concretization of the variable $x_j$ in the CTS $S$, $1\le j \le n$, consists in the assignment of a constant value to this variable:
$x_j\equiv 0$ or $x_j\equiv 1$, which entails removal from the CTS of all lines with an inverse value of $x_j$, with the subsequent clearing of the structure. The formal notation for this operation: $S(x_j \to 0)$ or $S(x_j \to 1)$. Thus, for $S_3$ and $S_4$ the following concretizations are possible:

\smallskip
$$S_3(x_3\to 1)$$
$$
\begin{array}{ccccccc}
x_1& x_2& x_3& x_4& x_5\\[5pt]
0&1&1&&\\
&1&1&0&\\
&&1&0&0\\
&&1&0&1&,
\end{array}
$$

\smallskip
$$S_4(x_5 \to 0)=\oslash .$$

Generalization of this unary operation for $n$ variables is natural: we consider an $n$-ary concretization as a unary operation applied $n$ times to $n$ variables. An example of denotation for
the generalized operation for $n=3$ is as follows: $S(x_1,\ x_2,\ x_5 \to 0,\ 1,\ 0)$.
Let us fix that the clearing procedure is automatically performed after the concretization and intersection operations.

{\it A substructure} $S'$ of the CTS $S$ is CTS composed of the subsets of the lines that are compatible at the adjacent tiers of $S$ (the notation for the operations: $S' \subseteq S$).
Note that a substructure, that is not empty, consists of $n-2$ tiers like any CTS.
   We define an {\it elementary} CTS as a CTS containing only one line at each tier. An elementary CTS corresponds to a single set of truth values and, accordingly, to a single SS. It means that CTS can be formed on the basis of any given system of binary sets (with interpretation of  0 and 1 as truth values). This purpose is attained by performing the union operation for the system of elementary CT structures corresponding to mentioned  sets.

Obviously, the general set of CT structures based on the {\it fixed permutation} of variables, with operations $\cup$ and $\cap$, forms a non-distributive lattice. An element of a  lattice---CTS is itself specifically structured and consists of the elementary CT structures. The set of elementary CT structures as operands of union operation is {\it not closed} in regard to this operation. Hence, the union of CT structures (not only elementary ones) is the CTS including, in general, besides CTS-operands, {\it superfluous} substructures, playing an essential part in the following analysis. Thus, in Section 4 we put into operation {\it a filtration procedure} for CTS, assigned to eliminate the superfluous substructure influence on the result of formula classification.

Let $S_1,\ S_2,\ \ldots ,\ S_q$ be the system of CT structures based on {\it different permutations} of variables (referred to as {\it discordant structures}). We define a $q$-ary operation of {\it unification} for the system as a special kind of a concretization performed in accordance with the next rules of joint transformation of the structures:

\hangindent=1.1cm
1)  If some variable $x_j\equiv 0$ or $x_j\equiv 1$ in at least one CTS $S_p$ ($1\le j\le n$, $1\le p\le q$), then all the lines containing inverse value of this variable have to be removed from all CT structures.

\hangindent=1.1cm
2)  If two variables $x_j$ and $x_r$ appear together (in any order) in compact triplets inside two or more CT structures, then the values combinations for these variables must be the same in all such structures. All the lines that are in contradiction with this constraint have to be removed from these structures.

\hangindent=1.1cm
3)  The clearing procedure accompanies each event of lines removal from the CT structures.

Herewith, variants appear when the unified system turns out to be an empty set:
\begin{itemize}
\item a certain tier in at least one CTS turns out to be empty;
\item there exists a ``conflict" of constant
values for a certain variable $x_j$ in at least two CT structures, i.e. $x_j\equiv 0$ in one structure and  $x_j\equiv 1$
in another structure;
\item at least one CTS is originally empty (a trivial case).
\end{itemize}

Thus, the unified CT structures can be either empty or non-empty only simultaneously.

\par\medskip

\textbf{3. Formula decomposition}

In general case it is necessary for 3-SAT problem resolution to decompose the initial formula $F$  using the operation: $F=F_1\wedge F_2\wedge  \cdots \wedge F_k$,  where $F_r$ , ${r=1,\ 2,\, \ldots,\ k,}\ k\le m$, is the formula suitable for CTF presentation based on the individual permutation of variables ${P_r=\langle x_{r_1},\ x_{r_2}, \ldots,\ x_{r_n}\rangle}$.

The decomposition requires a polynomial procedure which consists of following points:
\begin{itemize}
\item grouping the lines of  $F$  with identical numbers of three
	non-empty columns;
\item	putting three non-empty columns including the symbols of the variables in each of $k$ obtained groups ($k$ matrices) into the places 1, 2, 3 with shifting the other columns; it causes fixation of  $k$ permutations of the variables as bases for  $k$  CTF.
\end{itemize}

So, the final  $k$  matrices are the ordinary CT formulas. Note that empty tiers are permitted in CTF, in contrast to CTS.

The described procedure comes to $k$-tuple survey of the lines of $nm$-matrix, hence the estimation of the complexity of a decomposition algorithm is $O(mnk)$. The suitable permutations are obtained by forming and not by enumeration; that results in elimination of exponential computation complexity.

The modernized algorithms can be based on different methods of assembling CTF out of matrices consisting of the first three columns of the CT formulas obtained by the previous algorithm; these columns are considered as tiers in lesser quantity of CTF.

Clearly, the parameter $k$ satisfies the condition  $\lceil w/(n-2)\rceil \le k \le m$, where  $w$  is the number of groups containing terms (the elementary disjunctions) with identical variables. For an ``ideal" formula  $F$,  $k = 1$; the extreme value $k = m$  relates to forming a separate permutation for each term of the initial formula. Note that we put aside possible methods of minimizing  $k$ as a non-principal point of the model realization.

Then we transform each CTF $F_r$ to CTS  $S_r$. Now the problem is reduced to the following one: ascertain the fact of existence (or absence) of {\it joint satisfying sets} (abbreviations: {\it JS sets} or {\it JSS}) for the system of discordant CT structures
$S_1,\ S_2,\, \dots, \ S_k$.  It is necessary to solve this new problem without a searching through the sets, coded in the CT structures, in order to avoid procedures of exponential complexity.

In order to illustrate theoretical aspects of the model realization (without restriction of the general analysis) we use, as an example, the initial tabular formula $F$ shown in Table 1.

The decomposition of $F$ was carried out with the use of assembling the tiers obtained by the procedure stated above. The resulting CT formulas based on three variable's permutations are presented in Table 2. Finally, CTF $\to$ CTS transformation described at Section 2 leads to the three CT structures:  $S_1$, $S_2$ and $S_3$ (Table 3).

\begin{table}[h]
\begin{tabular}{|c|c|c|c|c|c|c|c|c|c|c|c|c|c|c|c|c|c|c|c|c|c|c|c|c|c|}
\multicolumn{26}{c}{\bf Table 1. Initial formula $F$}\\
\multicolumn{26}{c}{}\\
\cline{1-8}
\cline{10-17}
\cline{19-26}
$a$&$b$&$c$&$d$&$e$&$f$&$g$&$h$&&$a$&$b$&$c$&$d$&$e$&$f$&$g$&$h$&&$a$&$b$&$c$&$d$&$e$&$f$&$g$&$h$\\
\hhline{|=|=|=|=|=|=|=|=|~|=|=|=|=|=|=|=|=|~|=|=|=|=|=|=|=|=|}
0&0&0&&&&&&&0&&1&&&&&0&&&&1&1&0&&&\\
\cline{1-8}                                                        \cline{10-17}
\cline{19-26}
0&1&0&&&&&&&1&&0&&&&&1&&&&1&0&&0&&\\
\cline{1-8}                                                        \cline{10-17}
\cline{19-26}
0&1&1&&&&&&&1&&1&&&&&1&&&&0&0&&0&&\\
\cline{1-8}                                                        \cline{10-17}
\cline{19-26}
1&0&0&&&&&&&&1&&&&&0&0&&&&1&&0&&&1\\
\cline{1-8}                                                        \cline{10-17}
\cline{19-26}
1&1&1&&&&&&&&0&&&&&1&0&&&&1&&0&&&0\\
\cline{1-8}                                                        \cline{10-17}
\cline{19-26}
0&1&&&0&&&&&&0&&&&&0&1&&&&&0&1&0&&\\
\cline{1-8}                                                        \cline{10-17}
\cline{19-26}
1&&&&1&0&&&&&1&&&&&0&1&&&&&1&0&1&&\\
\cline{1-8}                                                        \cline{10-17}
\cline{19-26}
1&&&&0&1&&&&&1&&&&&1&1&&&&&&0&0&0&\\
\cline{1-8}                                                        \cline{10-17}
\cline{19-26}
0&&0&&&0&&&&&0&&&0&&0&&&&&&&1&0&0&\\
\cline{1-8}                                                        \cline{10-17}
\cline{19-26}
0&&0&&&1&&&&&1&&&1&&1&&&&&&&1&1&1&\\
\cline{1-8}                                                        \cline{10-17}
\cline{19-26}
1&&1&&&0&&&&&0&&&0&&1&&&&&&&&0&1&0\\
\cline{1-8}                                                        \cline{10-17}
\cline{19-26}
1&&0&&&1&&&&&1&&&1&&&1&&&&&&&1&0&1\\
\cline{1-8}                                                        \cline{10-17}
\cline{19-26}
1&&&0&&0&&&&&0&&&0&&&0&&&&1&0&0&&&\\
\cline{1-8}                                                        \cline{10-17}
\cline{19-26}
0&&&0&&1&&&&&1&&&0&&0&&&0&&&1&&1&&\\
\cline{1-8}                                                        \cline{10-17}
\cline{19-26}
1&&&0&&1&&&&&&0&0&0&&&\\
\cline{1-8}                                                        \cline{10-17}

\end{tabular}
\end{table}

\newpage

\begin{table}[h]
               \begin{tabular}{|c|c|c|c|c|c|c|c|c|c|c|c|c|c|c|c|c|c|c|c|c|c|c|c|c|c|}
\multicolumn{26}{c}{\bf Table 2. CT formulas}\\
\multicolumn{26}{c}{$F_1$\hspace{5cm}$F_2$\hspace{5cm}$F_3$}\\
\cline{1-8}
\cline{10-17}
\cline{19-26}
$a$&$b$&$c$&$d$&$e$&$f$&$g$&$h$&&$a$&$b$&$c$&$d$&$e$&$f$&$g$&$h$&&$a$&$b$&$c$&$d$&$e$&$f$&$g$&$h$\\
\hhline{|=|=|=|=|=|=|=|=|~|=|=|=|=|=|=|=|=|~|=|=|=|=|=|=|=|=|}
0&0&0&&&&&&&0&0&1&&&&&&&0&0&1&&&&&\\
\cline{1-8}                                                        \cline{10-17}
\cline{19-26}
0&1&0&&&&&&&0&1&0&&&&&&&0&1&0&&&&&\\
\cline{1-8}                                                        \cline{10-17}
\cline{19-26}
0&1&1&&&&&&&1&0&0&&&&&&&0&1&1&&&&&\\
\cline{1-8}                                                        \cline{10-17}
\cline{19-26}
1&0&0&&&&&&&1&0&1&&&&&&&1&1&0&&&&&\\
\cline{1-8}                                                        \cline{10-17}
\cline{19-26}
1&1&1&&&&&&&1&1&1&&&&&&&&0&0&0&&&&\\
\cline{1-8}                                                        \cline{10-17}
\cline{19-26}
&&0&0&0&&&&&&0&0&0&&&&&&&0&1&1&&&&\\
\cline{1-8}                                                       \cline{10-17}
\cline{19-26}
&&1&0&0&&&&&&1&1&1&&&&&&&&0&1&0&&&\\
\cline{1-8}                                                        \cline{10-17}
\cline{19-26}
&&1&1&0&&&&&&1&0&0&&&&&&&&1&0&1&&&\\
\cline{1-8}                                                        \cline{10-17}
\cline{19-26}
&&&0&1&0&&&&&&1&0&0&&&&&&&1&1&1&&&\\
\cline{1-8}                                                        \cline{10-17}
\cline{19-26}
&&&1&0&1&&&&&&&1&1&0&&&&&&&1&1&0&&\\
\cline{1-8}                                                        \cline{10-17}
\cline{19-26}
&&&&0&0&0&&&&&&0&1&1&&&&&&&1&0&0&&\\
\cline{1-8}                                                        \cline{10-17}
\cline{19-26}
&&&&1&0&0&&&&&&&0&0&0&&&&&&&0&0&0&\\
\cline{1-8}                                                        \cline{10-17}
\cline{19-26}
&&&&1&1&1&&&&&&&0&1&0&&&&&&&1&1&1&\\
\cline{1-8}                                                        \cline{10-17}
\cline{19-26}
&&&&&0&1&0&&&&&&1&0&1&&&&&&&&0&1&0\\
\cline{1-8}                                                        \cline{10-17}
\cline{19-26}
&&&&&1&0&1&&&&&&1&1&0&&&&&&&&1&1&1\\
\cline{1-8}
\cline{10-17}
\cline{19-26}
\multicolumn{8}{}{}&&&&&&&0&1&0\\

\cline{10-17}

\multicolumn{8}{}{}&&&&&&&0&0&0\\

\cline{10-17}

\end{tabular}
\end{table}

\medskip
\textbf {4. Solution of the problem of JSS for two CT structures}

\smallskip
The resolution of 3-SAT problem for the formula reduced to two CT structures is a clue to the solution of the general problem. Let  $S_1$  and  $S_2$  be the two CT structures based on different permutations of variables (we use the structures from Table 3). The primary stage of CTS processing consists in unification operation for $S_1$ and $S_2$. This operation simplifies the CTS-operands by removing some of the lines that do not belong to JS sets, but preserves JS sets (if there exist any) in accordance with the operation rules. By this reason, we do not change notation for the unified CT structures (Table 4).

Theoretically, for the initial CT structures  $S_1$  and  $S_2$  there exist two {\em optimal} CT structures  $S_1^0$  and  $S_2^0$, each formed as the union of JS sets. In general, $S_1^0$ and  $S_2^0$  are primarily unknown but potentially existing mathematical objects. These structures, according to the construction rules, are empty if there are no JS sets for $S_1$  and  $S_2$ . In what follows we use a concept of optimal structures with some evident properties for the foundation of Theorem~1 and the main algorithm.

Let $S_1$ be a {\it basic structure}; we fix for it the initial numeration of variables: $x_1, \,x_2$, \,\ldots ,\,~$x_n$
($a, \,b$, \,\ldots ,\ $h$, in the presented example). Then we state that an alternative form for CTS representation is a graph  (a {\it basic graph} for  $S_1$). A graph for the CTS is a structured graph with the vertices located at  $n-2$  tiers in accordance with the location of the lines at CTS tiers; each vertex corresponds to some line and is marked with the triplet values of the line. Such marking together with the tier's number identify any vertex as an element of two-dimensional array; this mode of identification will be useful in following description of a hyperstructure.

\bigskip

\begin{table}[tbp]
\begin{tabular}{|c|c|c|c|c|c|c|c|c|c|c|c|c|c|c|c|c|c|c|c|c|c|c|c|c|c|}
\multicolumn{26}{c}{\bf Table 3. CT structures}\\
\multicolumn{26}{c}{$S_1$\hspace{5cm}$S_2$\hspace{5cm}$S_3$}\\
\cline{1-8}
\cline{10-17}
\cline{19-26}
$a$&$b$&$c$&$d$&$e$&$f$&$g$&$h$&&$h$&$g$&$b$&$e$&$a$&$f$&$c$&$d$&&
$d$&$f$&$a$&$c$&$h$&$e$&$b$&$g$\\
\hhline{|=|=|=|=|=|=|=|=|~|=|=|=|=|=|=|=|=|~|=|=|=|=|=|=|=|=|}
0&0&1&&&&&& &0&0&0&&&&&& &0&0&0&&&&&\\
\cline{1-8}                                                        \cline{10-17}
\cline{19-26}
1&0&1&&&&&& &0&1&1&&&&&& &1&0&0&&&&&\\
\cline{1-8}                                                        \cline{10-17}
\cline{19-26}
1&1&0&&&&&& &1&1&0&&&&&& &1&0&1&&&&&\\
\cline{1-8}                                                        \cline{10-17}
\cline{19-26}
&0&1&0&&&&& &&0&0&1&&&&& &1&1&1&&&&&\\
\cline{1-8}                                                        \cline{10-17}
\cline{19-26}
&0&1&1&&&&& &&1&0&1&&&&& &&0&0&1&&&&\\
\cline{1-8}                                                        \cline{10-17}
\cline{19-26}
&1&0&0&&&&& &&1&1&0&&&&& &&0&1&0&&&&\\
\cline{1-8}                                                        \cline{10-17}
\cline{19-26}
&1&0&1&&&&& &&&0&1&0&&&& &&1&1&0&&&&\\
\cline{1-8}                                                        \cline{10-17}
\cline{19-26}
&&0&0&1&&&& &&&0&1&1&&&& &&1&1&1&&&&\\
\cline{1-8}                                                        \cline{10-17}
\cline{19-26}
&&0&1&0&&&& &&&1&0&1&&&& &&&0&1&1&&&\\
\cline{1-8}                                                        \cline{10-17}
\cline{19-26}
&&0&1&1&&&& &&&&1&0&0&&& &&&1&0&0&&&\\
\cline{1-8}                                                        \cline{10-17}
\cline{19-26}
&&1&0&1&&&& &&&&1&0&1&&& &&&1&1&0&&&\\
\cline{1-8}                                                        \cline{10-17}
\cline{19-26}
&&1&1&1&&&& &&&&0&1&0&&& &&&&1&1&1&&\\
\cline{1-8}                                                        \cline{10-17}
\cline{19-26}
&&&0&1&1&&& &&&&1&1&1&&& &&&&0&0&0&&\\
\cline{1-8}                                                        \cline{10-17}
\cline{19-26}
&&&1&0&0&&& &&&&&0&0&1&& &&&&0&0&1&&\\
\cline{1-8}                                                        \cline{10-17}
\cline{19-26}
&&&1&1&0&&& &&&&&0&1&1&& &&&&1&0&1&&\\
\cline{1-8}                                                        \cline{10-17}
\cline{19-26}
&&&1&1&1&&& &&&&&1&0&0&& &&&&&1&1&0&\\
\cline{1-8}                                                        \cline{10-17}
\cline{19-26}
&&&&0&0&1&& &&&&&1&1&1&& &&&&&0&0&1&\\
\cline{1-8}                                                        \cline{10-17}
\cline{19-26}
&&&&1&0&1&& &&&&&&0&1&1& &&&&&0&1&0&\\
\cline{1-8}                                                        \cline{10-17}
\cline{19-26}
&&&&1&1&0&& &&&&&&0&0&1& &&&&&0&1&1&\\
\cline{1-8}                                                        \cline{10-17}
\cline{19-26}
&&&&&0&1&1& &&&&&&1&1&0& &&&&&&1&0&1\\
\cline{1-8}                                                        \cline{10-17}
\cline{19-26}
&&&&&1&0&0& &&&&&&1&1&1& &&&&&&0&1&1\\
\cline{1-8}                                                        \cline{10-17}
\cline{19-26}
\multicolumn{17}{}{}   &&&&&&&1&0&0\\
\cline{19-26}
\multicolumn{17}{}{}   &&&&&&&1&1&0\\
\cline{19-26}
\end{tabular}
\end{table}

\begin{table}[p]
\centering
\begin{tabular}{|c|c|c|c|c|c|c|c|c|c|c|c|c|c|c|c|c|}
\multicolumn{17}{c}{\bf Table 4. Unified CT structures \,$S_1$ and  $S_2$}\\
\multicolumn{17}{c}{$S_1$\hspace{5cm}$S_2$}\\
\cline{1-8}
\cline{10-17}
$a$&$b$&$c$&$d$&$e$&$f$&$g$&$h$&&$h$&$g$&$b$&$e$&$a$&$f$&$c$&$d$\\
\hhline{|=|=|=|=|=|=|=|=|~|=|=|=|=|=|=|=|=|}
0&0&1&&&&&& &0&0&0&&&&&\\
\cline{1-8}                                                        \cline{10-17}
1&0&1&&&&&& &1&1&0&&&&&\\
\cline{1-8}                                                        \cline{10-17}
&0&1&0&&&&& &&0&0&1&&&&\\
\cline{1-8}                                                        \cline{10-17}
&0&1&1&&&&& &&1&0&1&&&&\\
\cline{1-8}                                                        \cline{10-17}
&&1&0&1&&&& &&&0&1&0&&&\\
\cline{1-8}                                                        \cline{10-17}
&&1&1&1&&&& &&&0&1&1&&&\\
\cline{1-8}                                                        \cline{10-17}
&&&0&1&1&&& &&&&1&0&0&&\\
\cline{1-8}                                                        \cline{10-17}
&&&1&1&0&&& &&&&1&0&1&&\\
\cline{1-8}                                                        \cline{10-17}
&&&1&1&1&&& &&&&1&1&1&&\\
\cline{1-8}                                                        \cline{10-17}
&&&&1&0&1&& &&&&&0&0&1&\\
\cline{1-8}                                                        \cline{10-17}
&&&&1&1&0&& &&&&&0&1&1&\\
\cline{1-8}                                                        \cline{10-17}
&&&&&0&1&1& &&&&&1&1&1&\\
\cline{1-8}                                                        \cline{10-17}
&&&&&1&0&0& &&&&&&0&1&1\\
\cline{1-8}                                                        \cline{10-17}
\multicolumn{8}{}{}    &&&&&&&1&1&0\\
                                                        \cline{10-17}
\multicolumn{8}{}{}    &&&&&&&1&1&1\\
                                                        \cline{10-17}
\end{tabular}
\end{table}

The edges of the graph correspond to pairs of lines adjoined at coinciding of two values at neighboring tiers. Each vertex is joined by one or two edges with vertices of the tier-predecessor and the tier-successor in accordance with the CTS construction. The JS sets are associated with {\it routes} in the basic graph; each route includes one distinct vertex of each tier and edges that join these vertices. We state an agreement that only such routes are the subjects of our consideration. The graph $G_1=(V,E)$ (Fig. 1) corresponds to CTS $S_1$ (Table~4). The determination of JSS for $S_1$ and $S_2$
denotes satisfiability of a formula $F'$ presented by the subset of the lines in Table 1 (the lines that served for forming  $S_1$  and  $S_2$  before the unification of these structures).

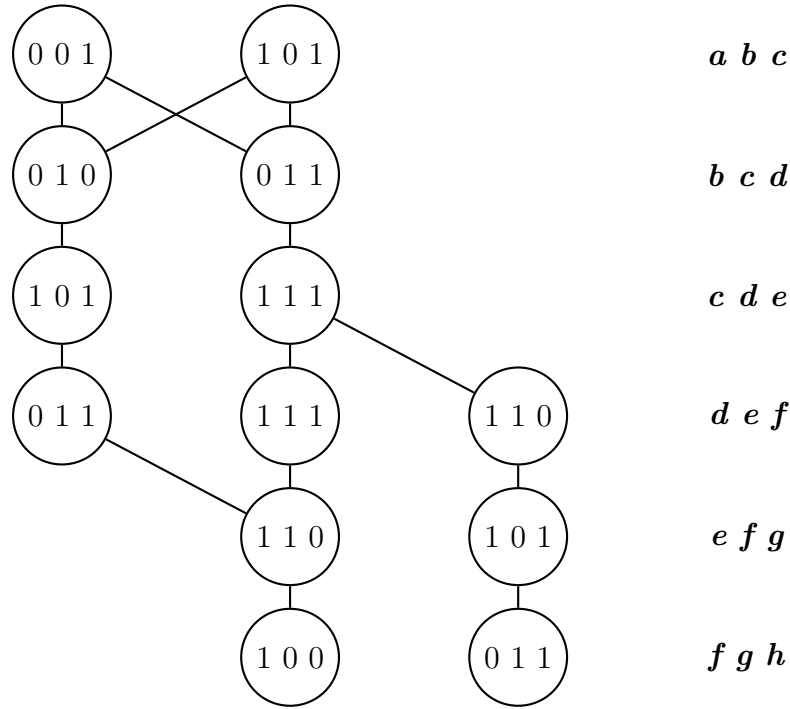
\begin{figure}[h]
\begin{center}
\begin{tikzpicture}
[
	xscale	= 1,	
	yscale	= 1,	
]

{
	\node (nJOneOne) [shape = circle, draw = black, thick] at (0,0) {0 0 1};
           \node (nJOneTwo) [shape = circle, draw = black, thick] at (3,0) {1 0 1};
	\node (nJTwoOne) [shape = circle, draw = black, thick] at (0,-1.6) {0 1 0};
           \node (nJTwoTwo) [shape = circle, draw = black, thick] at (3,-1.6) {0 1 1};
	\node (nJThreeOne) [shape = circle, draw = black, thick] at (0,-3.2) {1 0 1};
           \node (nJThreeTwo) [shape = circle, draw = black, thick] at (3,-3.2) {1 1 1};
	\node (nJFourOne) [shape = circle, draw = black, thick] at (0,-4.8) {0 1 1};
           \node (nJFourTwo) [shape = circle, draw = black, thick] at (3,-4.8) {1 1 1};
	\node (nJFourThree) [shape = circle, draw = black, thick] at (6,-4.8) {1 1 0};
	\node (nJFiveOne) [shape = circle, draw = black, thick] at (3,-6.4) {1 1 0};
           \node (nJFiveTwo) [shape = circle, draw = black, thick] at (6,-6.4) {1 0 1};
	\node (nJSixOne) [shape = circle, draw = black, thick] at (3,-8.0) {1 0 0};
           \node (nJSixTwo) [shape = circle, draw = black, thick] at (6,-8.0) {0 1 1};

	\draw (9,0) node {\bf\em a b c};
	\draw (9,-1.6) node {\bf\em b c d};
	\draw (9,-3.2) node {\bf\em c d e};
	\draw (9,-4.8) node {\bf\em d e f};
	\draw (9,-6.4) node {\bf\em e f g};
	\draw (9,-8.0) node {\bf\em f g h};

	\draw [thick, -] (nJOneOne) -- (nJTwoTwo);
	\draw [thick, -] (nJTwoOne) -- (nJOneTwo);
	\draw [thick, -] (nJOneOne) -- (nJTwoOne);
	\draw [thick, -] (nJOneTwo) -- (nJTwoTwo);
	\draw [thick, -] (nJTwoOne) -- (nJThreeOne);
	\draw [thick, -] (nJTwoTwo) -- (nJThreeTwo);
	\draw [thick, -] (nJThreeOne) -- (nJFourOne);
	\draw [thick, -] (nJThreeTwo) -- (nJFourTwo);
	\draw [thick, -] (nJThreeTwo) -- (nJFourThree);
	\draw [thick, -] (nJFourOne) -- (nJFiveOne);
	\draw [thick, -] (nJFourTwo) -- (nJFiveOne);
	\draw [thick, -] (nJFourThree) -- (nJFiveTwo);
	\draw [thick, -] (nJFiveOne) -- (nJSixOne);
	\draw [thick, -] (nJFiveTwo) -- (nJSixTwo);
}

\end{tikzpicture}
\end{center}
\caption{Basic graph}
\end{figure}

{\bf \underline {Hyperstructure}}.
 Further on, we put into operation a new structural object---a {\it  hyperstructure} (HS). Let $G_2$  be a {\it copy} of the basic graph $G_1$ preserving the notation of vertices and edges fixed for $G_1$. The graph  $G_2$   will be a {\it skeleton} of the hyperstructure  $\Gamma$, and its elements (vertices and edges) will be called {\it copies} of {\it same-name} elements of the basic graph (BG). The hyperstructure  $\Gamma$  itself is a structured graph constructed by assignment to each vertex  $v_j$  of  $G_2$  a {\it substructure-vertex} $\pi_j\subseteq S_2$ formed by means of concretization of the three variables with the values pointed in the same-name vertex of the basic graph, accompanied by a {\it filtration} procedure (see below). Besides, a {\it substructure-edge}  $\pi_{jk}=\pi_j \cap \pi_k$   is  assigned  to  each  edge $(v_j,\ v_k)$.  Thus, the hyperstructure is a graph that preserves the topology and tiers of the basic graph but contains structured vertices and edges. Note that we do not use a full numeration for distinct vertices and edges in HS (it is not required in the further analysis because, as it will be stated, no binary operations are executed for substructures located at the same tier) and the only subscript at any vertex name will denote the tier number.

If some substructure-vertices or substructure-edges assigned to elements of HS turn out to be empty (because of concretization and filtration procedures), the same-name elements (vertices and edges) are to be removed from the graphs $G_1$ and $G_2$; removing any vertex involves removing all the incident edges. Moreover, according to the tier partition of BG and HS, no vertex at any tier can exist without adjacent vertices situated at both adjacent tiers. Lastly, if any tier of HS turns out to be empty, the whole HS is declared an empty set. Note that any intersections of substructure-vertices belonging to the same tier are empty, which is direct consequence of the tier construction rules.

{\bf \underline {Effective procedure for HS forming}}.
 The strict algorithmic formalization of HS forming principles leads to an {\it effective procedure} (EP)---the main part of the formula classification algorithm. We define now several important operations and HS components that will serve for the detailed definition of HS.

A substructure that is formed by means of intersection of $j$  substructures located respectively at the tiers 1, 2, \ \dots ,\ $j$\ $(1 \le j < n-3)$, is said to be a {\it $j$-intersection} of substructures in HS. Non-empty $j$-intersection can exist only if each two vertices of adjacent tiers, corresponding to mentioned substructures, are joined by an edge.

We define a {\it projection of the $r${\rm th} tier} onto the substructure  $\pi_j$ of the $j$th tier ($j~>~r$). The
operation consists of two parts: ({\bf a}) calculation of the intersections of  $\pi_j$ with each substructure of the $r$th tier; ({\bf b}) calculation of the union of these intersections. The resulting substructure replaces $\pi_j$.

Suppose that the tiers 1, 2, \ \dots ,\ $j$\ $(1 \le j < n-3, n \ge 4)$ have been formed in HS. Then we consider the principle of forming the substructures of  $(j+1)$th tier using the following notation:

$v_j$ 	\hspace{1.8cm} vertex of  the $j$th tier in HS;

$\pi_j$	\hspace{1.8cm} substructure-vertex assigned to  $v_j$;

($v_j$, $v_{j+1}$) \hspace{0.5cm} some selected edge incident to the vertices of the  $j$th  and  $(j+1)$th  tiers;

$\pi_{j,j+1}$ \hspace{1.2cm}	substructure-edge assigned to ($v_j$, $v_{j+1}$);

$x_{j+3}$  \hspace{1.4cm} variable pointed in BG for the vertex  $v_{j+1}$;

$\beta$ \hspace{1.9cm}	value of  $x_{j+3}$  fixed for the vertex $v_{j+1}$  in BG, $\beta \in \{0, 1\}$.

\medskip
A {\it shift} of the substructure-vertex  $\pi_j$  along the edge $(v_j,\,v_{j+1})$ with simultaneous calculation of the substructure-edge $\pi_{j,j+1}$ includes the concretization   $\pi_j(x_{j+3}\to \beta)$ and (if  $j \ge 2$) the successive projections of the tiers 1, 2,\, \dots ,\ $j-1$ onto the substructure received as a result of the concretization. If $j=1$, $\pi_{j,j+1}=\pi_j(x_{j+3}\to \beta)$. Strictly, the shift for $j \ge 2$ is described by iterations: $$\pi_j(\beta,0)=\pi_j(x_{j+3}\to \beta),$$
$$
\pi_j(\beta,s)=(\pi_s^1\cap \pi_j(\beta,s-1))\cup \ldots\cup (\pi_s^{k_s}\cap \pi_j(\beta,s-1)),\eqno (1)
$$
$$s=1,2, \ \dots ,\ j-1.$$

Here $\pi_j(\beta,s)$ is the result of projection of the $s$th tier onto the substructure $\pi_j(\beta,s-1)$; the upper index at $\pi_s$  is the number of the substructure that belongs to the $s$th tier $(s<j)$. In~(1) $\pi_j(\beta,s-1)$ cannot be factored out because the lattice of CT structures is non-distributive; the value $s-1=0$,
initiating the iterations when $s=1$, denotes no tier number.

By means of (1), $\pi_j(\beta,1), \pi_j(\beta,2), \ \dots ,\ \pi_j(\beta, j-1)$ are calculated successively. The last substructure  $\pi_j(\beta, j-1)$ is the objective substructure-edge assigned to the edge $(v_j,\,v_{j+1})$, i.e. $\pi_{j,j+1}=\pi_j(\beta, j-1)$.
The substructure $\pi_j$ after the calculation of $\pi_{j,j+1}$ is to be restored to the initial form.

If $v_j$ is the only vertex at the $j$th tier adjacent to $v_{j+1}$, then the substructure-vertex $\pi_{j+1} = \pi_{j,j+1}$;
if there exist two edges:  $(v_j,\,v_{j+1})$ and $(v_j',\,v_{j+1})$, then the substructure-vertex $\pi_{j+1}$ is calculated as a
union of two corresponding substructure-edges. The possible appearance of empty substructures in the calculation process involves the removal actions described above.

{\bf REMARK 1}. The succession of tier projection operations (1) onto the substructure  $\pi_j(x_{j+3}\to \beta), \beta\in \{0,1\}$,
{\it separates} from  $\pi_j$ a substructure $\pi_j(\beta)$ that preserves all  $j$-intersections with $x_{j+3}=\beta$, existing in $\pi_j$ (according to the definition of  $j$-intersection). We will call this succession of operations a {\it filtration} procedure for the substructure $\pi_j(x_{j+3}\to \beta)$. The aim of filtration is to prevent formation of substructure-vertices that are composed {\em only} of {\it superfluous} substructures, i.e., of substructures that are not $j$-intersections.

Thus, shifting the substructure-vertex $\pi_j$ along the edge ($v_j$,  $v_{j+1}$) is performed at the stage of HS formation and involves the filtration mechanism.

{\bf REMARK 2}. Clearly, $\pi_j(x_{j+3}\to \beta)\cap \pi^* = \pi_j\cap \pi^*(x_{j+3}\to \beta)$,  where $\pi^*$ is any CTS. It means that two next procedures calculate the same substructure $\pi_j(\beta)$: ({\bf a}) filtration of the substructure $\pi_j(x_{j+3}\to \beta)$; ({\bf b}) projection of the tiers 1, 2, \ \dots ,\ $j-1$ onto the substructure $\pi_j$ in a {\it special case} of identity $x_{j+3}\equiv \beta$ in all substructures that discover a non-empty intersection with $\pi_j$.

Thus, the substructure-edge $\pi_{j,j+1}$, as a product of the filtration procedure, is formed as a union of  $j$-intersections and, correspondingly, the part of the substructure-vertex $\pi_{j+1}$ that is generated by $\pi_{j,j+1}$ (or $\pi_{j+1}$ completely, if the vertex $v_{j+1}$ has the only adjacent vertex at the $j$th  tier) is formed as a union of $(j+1)$-intersections.

Note that, in accordance with the properties of non-distributive lattice formed by CTS sets, a union of  $j$-intersections is, in general, a substructure that includes, besides of union operands, superfluous CT substructures which are compositions of compact triplets from different $j$-intersections. Yet, owing to the filtration procedure, the shift for the superfluous substructures is possible only jointly with $j$-intersections (not autonomously). It will be clear from the further analysis (strictly, from Theorems 1 and 2) that forming each substructure of the $j$th tier as a union of  $j$-intersections is a decisive factor of the model efficiency.

\medskip
Now, after the given detailed definitions, we can enumerate the items of HS forming:

\hangindent=1.1cm
1 \  Forming the substructure-vertices of HS first tier by means of concretization in CTS $S_2$ of variables that have been fixed in the first compact triplet of BG; concretization is executed with the values pointed for the same-name vertices of BG and HS. If the first tier is absent (all substructures are empty), then the early termination of the EP follows ($\Gamma=\oslash$), else the transition to item 2 that describes iterative process is carried out.

\hangindent=1.1cm
2 \  Performing for $j=1, 2, \,\dots,\ n-3$ the next two items (on condition that the current completely formed tier is not empty, as it becomes clear from i. 2\  b).

\hangindent=1.5cm
2\ a)  Performing  the  shifts  of  the substructure-vertices  of  the  $j$th  tier along  the edges  showed in the basic graph (connecting the vertices of  the  $j$th  and  $(j+1)$th  tiers), with calculation of the substructure-edges assigned to these edges.

\hangindent=1.5cm
2\ b)  Forming the substructure-vertices of the $(j+1)$th  tier using the substructure-edges, defined in i. 2\ a. If the $(j+1)$th tier is empty, then the procedure terminates with the result $\Gamma=\oslash$. If the $(n-2)$th tier is not empty, then the procedure terminates with the result~$\Gamma\ne \oslash$.

\hangindent=1.1cm
3 \  Correction of BG in the course of HS skeleton modification, in order to preserve  $G_1$  and  $G_2$  isomorphism.

It is obvious that any substructure $\pi_j$ of the $j$th tier ($j\ge   2$), according to the filtration procedure, has a non-empty intersection with at least one substructure of each tier associated with numeration $j=1, 2, \,\dots,\ j-1$.

The HS for  $S_1$  and  $S_2$ CT structures defined by Table 3 is shown in Fig. 2; the identical substructures assigned to adjacent vertices of skeleton at the tiers 5 and 6 are not duplicated.

{\bf \underline {Routes in HS}}. Let us consider in general the HS that is completely formed on the basis of $S_1$ and $S_2$ CT structures ($\Gamma\ne \oslash$). We will call an {\it HS route} a sequence of vertices and edges in the HS skeleton (a sequence including one distinct vertex of each tier and the edges that join these vertices) on condition that the $(n-2)$-intersection of substructure-vertices assigned to vertices of a sequence is not empty. Note that any HS route is a copy of some route in BG, as the latter has been defined.

{\bf STATEMENT}. The mapping of the set of HS routes onto the set of JSS's for $S_1$ and $S_2$ is a bijection.

{\bf PROOF}. Let some distinct HS route $\mu $ be a route defined by the vertices enumeration: $\mu = (v_1, v_2, \ \dots ,\ v_{n-2})$.
The non-empty intersection $\pi_{\mu} = \pi_1\cap \pi_2\cap \,\ldots \, \cap \pi_{n-2}$ of corresponding
substructure-vertices separates from $S_2$ the elementary CTS (the satisfying set for $S_2$) because of concretization of all variables in $(n-2)$-intersection. The values of the fixed variables in $(n-2)$-intersection coincide with the values that are fixed for corresponding variables in the same-name route $(v_1, v_2, \ \dots ,\ v_{n-2})$ in BG, so the latter route determines the elementary CTS as a part of $S_1$ and as the satisfying set for $S_1$. Consequently, both satisfying sets represent a JSS.

Conversely, each JSS in the form of the elementary CTS is a part of all substructure-vertices that are associated with the unique HS route strictly determined by variable values common for  $S_1$ and  $S_2$. $\Box$

For the HS (Fig. 2) we discover, using the lower tier, five routes and, correspondingly, five JS sets:
$$
\begin{array}{cccccccc}
h&g&b&e&a&f&c&d\\[5pt]
0&0&0&1&0&1&1&0\\
0&0&0&1&0&1&1&1\\
0&0&0&1&1&1&1&0\\
0&0&0&1&1&1&1&1\\
1&1&0&1&0&0&1&1\\
\end{array}
$$

As is clear from the STATEMENT, {\it the existence of non-empty HS} \ is a necessary condition for an existence of JSS for two CT structures.

The way to proof the sufficiency of this condition is based on the concept of {\it optimal hyperstructure}. An optimal HS is a product of EP applied to the optimal CT structures $S_1^0$  and  $S_2^0$,
defined at the beginning of this section. An optimal HS, like optimal CT structures, is a potentially existing object that is not empty if and only if JS sets exist for the initial CT structures
$S_1$ and $S_2$. The crucial feature of the optimal HS is that each substructure-vertex at any tier,
according to the optimal CT structures definition and the HS construction rules, is a union of JS sets, viz. a union of $(n-2)$-intersections.

\begin{figure}[tbp]
\begin{center}
\begin{tikzpicture}[node distance=0.2mm]
[
	xscale	= 1,	
	yscale	= 1,	
]

{
         \small 


	\node (nJOneOne) {
           
		{\renewcommand{\arraystretch}{0.5}
		\begin{tabular}{@{}*{3}{c@{\hspace{0.5mm}}}}
			{\bf \em a}&{\bf \em b}&{\bf \em c}\\[0.5mm]0&0&1
		\end{tabular}}

};

	\node (nJOneTwo) [right=3cm of nJOneOne] {
           
		{\renewcommand{\arraystretch}{0.5}
		\begin{tabular}{@{}*{3}{c@{\hspace{0.5mm}}}}
			{\bf \em a}&{\bf \em b}&{\bf \em c}\\[0.5mm]1&0&1
		\end{tabular}}

};

	\node (nJTwoOne) [below=3.3cm of nJOneOne] {
           
		{\renewcommand{\arraystretch}{0.5}
		\begin{tabular}{@{}*{3}{c@{\hspace{0.5mm}}}}
			{\bf \em b}&{\bf \em c}&{\bf \em d}\\[0.5mm]0&1&0
		\end{tabular}}

};

	\node (nJTwoTwo) [below=3.3cm of nJOneTwo] {
           
		{\renewcommand{\arraystretch}{0.5}
		\begin{tabular}{@{}*{3}{c@{\hspace{0.5mm}}}}
			{\bf \em b}&{\bf \em c}&{\bf \em d}\\[0.5mm]0&1&1
		\end{tabular}}

};

	\node (nJThreeOne) [below=3.2cm of nJTwoOne] {
           
		{\renewcommand{\arraystretch}{0.5}
		\begin{tabular}{@{}*{3}{c@{\hspace{0.5mm}}}}
			{\bf \em c}&{\bf \em d}&{\bf \em e}\\[0.5mm]1&0&1
		\end{tabular}}

};

	\node (nJThreeTwo) [below=3.2cm of nJTwoTwo] {
           
		{\renewcommand{\arraystretch}{0.5}
		\begin{tabular}{@{}*{3}{c@{\hspace{0.5mm}}}}
			{\bf \em c}&{\bf \em d}&{\bf \em e}\\[0.5mm]1&1&1
		\end{tabular}}

};

	\node (nJFourOne) [below=3.2cm of nJThreeOne] {
           
		{\renewcommand{\arraystretch}{0.5}
		\begin{tabular}{@{}*{3}{c@{\hspace{0.5mm}}}}
			{\bf \em d}&{\bf \em e}&{\bf \em f}\\[0.5mm]0&1&1
		\end{tabular}}

};

	\node (nJFourTwo) [below=3.2cm of nJThreeTwo] {
           
		{\renewcommand{\arraystretch}{0.5}
		\begin{tabular}{@{}*{3}{c@{\hspace{0.5mm}}}}
			{\bf \em d}&{\bf \em e}&{\bf \em f}\\[0.5mm]1&1&1
		\end{tabular}}

};

	\node (nJFourThree) [right=3cm of nJFourTwo] {
           
		{\renewcommand{\arraystretch}{0.5}
		\begin{tabular}{@{}*{3}{c@{\hspace{0.5mm}}}}
			{\bf \em d}&{\bf \em e}&{\bf \em f}\\[0.5mm]1&1&0
		\end{tabular}}

};

	\node (nJFiveOne) [below=2cm of nJFourTwo] {
           
		{\renewcommand{\arraystretch}{0.5}
		\begin{tabular}{@{}*{3}{c@{\hspace{0.5mm}}}}
			{\bf \em e}&{\bf \em f}&{\bf \em g}\\[0.5mm]1&1&0
		\end{tabular}}

};

	\node (nJFiveTwo) [below=2cm of nJFourThree] {
           
		{\renewcommand{\arraystretch}{0.5}
		\begin{tabular}{@{}*{3}{c@{\hspace{0.5mm}}}}
			{\bf \em e}&{\bf \em f}&{\bf \em d}\\[0.5mm]1&0&1
		\end{tabular}}

};
	\node (nJSixOne) [below=0.4cm of nJFiveOne] {
           
		{\renewcommand{\arraystretch}{0.5}
		\begin{tabular}{@{}*{3}{c@{\hspace{0.5mm}}}}
			{\bf \em f}&{\bf \em g}&{\bf \em h}\\[0.5mm]1&0&0
		\end{tabular}}

};

	\node (nJSixTwo) [below=0.4cm of nJFiveTwo] {
           
		{\renewcommand{\arraystretch}{0.5}
		\begin{tabular}{@{}*{3}{c@{\hspace{0.5mm}}}}
			{\bf \em f}&{\bf \em g}&{\bf \em h}\\[0.5mm]0&1&1
		\end{tabular}}

};

\begin{scope}[auto,bend left]
	\draw [thick, -] (nJOneOne) to (nJTwoTwo);
	\draw [thick, -] (nJTwoOne) to (nJOneTwo);
	\draw [thick, -] (nJOneOne) -- (nJTwoOne);
	\draw [thick, -] (nJOneTwo) -- (nJTwoTwo);
	\draw [thick, -] (nJTwoOne) -- (nJThreeOne);
	\draw [thick, -] (nJTwoTwo) -- (nJThreeTwo);
	\draw [thick, -] (nJThreeOne) -- (nJFourOne);
	\draw [thick, -] (nJThreeTwo) -- (nJFourTwo);
	\draw [thick, -] (nJThreeTwo) to (nJFourThree);
	\draw [thick, -] (nJFourOne) to (nJFiveOne);
	\draw [thick, -] (nJFourTwo) -- (nJFiveOne);
	\draw [thick, -] (nJFourThree) -- (nJFiveTwo);
	\draw [thick, -] (nJFiveOne) -- (nJSixOne);
	\draw [thick, -] (nJFiveTwo) -- (nJSixTwo);
\end{scope}

	\node (nHOneOne) [left=of nJOneOne] {
		{\renewcommand{\arraystretch}{0.5}
		\begin{tabular}{@{}*{8}{c@{\hspace{0.5mm}}}}
			{\bf \em h}&{\bf \em g}&{\bf \em b}&{\bf \em e}&{\bf \em a}&{\bf \em f}&{\bf \em c}&{\bf \em d}\\[0.5mm]
0&0&0\\
1&1&0\\
&0&0&1\\
&1&0&1\\
&&0&1&0\\
&&&1&0&0\\
&&&1&0&1\\
&&&&0&0&1\\
&&&&0&1&1\\
&&&&&0&1&1\\
&&&&&1&1&0\\
&&&&&1&1&1\\
		\end{tabular}}
};

	\node (nHOneTwo) [right=of nJOneTwo] {
		{\renewcommand{\arraystretch}{0.5}
		\begin{tabular}{@{}*{8}{c@{\hspace{0.5mm}}}}
			{\bf \em h}&{\bf \em g}&{\bf \em b}&{\bf \em e}&{\bf \em a}&{\bf \em f}&{\bf \em c}&{\bf \em d}\\[0.5mm]
0&0&0\\
1&1&0\\
&0&0&1\\
&1&0&1\\
&&0&1&1\\
&&&1&1&1\\
&&&&1&1&1\\
&&&&&1&1&0\\
&&&&&1&1&1\\
		\end{tabular}}

	};
	\node (nHTwoOne) [left=of nJTwoOne] {
		{\renewcommand{\arraystretch}{0.5}
		\begin{tabular}{@{}*{8}{c@{\hspace{0.5mm}}}}
			{\bf \em h}&{\bf \em g}&{\bf \em b}&{\bf \em e}&{\bf \em a}&{\bf \em f}&{\bf \em c}&{\bf \em d}\\[0.5mm]
0&0&0\\
1&1&0\\
&0&0&1\\
&1&0&1\\
&&0&1&0\\
&&0&1&1\\
&&&1&0&1\\
&&&1&1&1\\
&&&&0&1&1\\
&&&&1&1&1\\
&&&&&1&1&0\\
		\end{tabular}}

};

	\node (nHTwoTwo) [right=of nJTwoTwo] {
		{\renewcommand{\arraystretch}{0.5}
		\begin{tabular}{@{}*{8}{c@{\hspace{0.5mm}}}}
			{\bf \em h}&{\bf \em g}&{\bf \em b}&{\bf \em e}&{\bf \em a}&{\bf \em f}&{\bf \em c}&{\bf \em d}\\[0.5mm]
0&0&0\\
1&1&0\\
&0&0&1\\
&1&0&1\\
&&0&1&0\\
&&0&1&1\\
&&&1&0&0\\
&&&1&0&1\\
&&&1&1&1\\
&&&&0&0&1\\
&&&&0&1&1\\
&&&&1&1&1\\
&&&&&0&1&1\\
&&&&&1&1&1\\

		\end{tabular}}

};

	\node (nHThreeOne) [left=of nJThreeOne] {
		{\renewcommand{\arraystretch}{0.5}
		\begin{tabular}{@{}*{8}{c@{\hspace{0.5mm}}}}
			{\bf \em h}&{\bf \em g}&{\bf \em b}&{\bf \em e}&{\bf \em a}&{\bf \em f}&{\bf \em c}&{\bf \em d}\\[0.5mm]
0&0&0\\
1&1&0\\
&0&0&1\\
&1&0&1\\
&&0&1&0\\
&&0&1&1\\
&&&1&0&1\\
&&&1&1&1\\
&&&&0&1&1\\
&&&&1&1&1\\
&&&&&1&1&0\\

		\end{tabular}}

};

	\node (nHThreeTwo) [left=of nJThreeTwo] {
		{\renewcommand{\arraystretch}{0.5}
		\begin{tabular}{@{}*{8}{c@{\hspace{0.5mm}}}}
			{\bf \em h}&{\bf \em g}&{\bf \em b}&{\bf \em e}&{\bf \em a}&{\bf \em f}&{\bf \em c}&{\bf \em d}\\[0.5mm]

0&0&0\\
1&1&0\\
&0&0&1\\
&1&0&1\\
&&0&1&0\\
&&0&1&1\\
&&&1&0&0\\
&&&1&0&1\\
&&&1&1&1\\
&&&&0&0&1\\
&&&&0&1&1\\
&&&&1&1&1\\
&&&&&0&1&1\\
&&&&&1&1&1\\

		\end{tabular}}

};
	\node (nHFourOne) [left=of nJFourOne] {
		{\renewcommand{\arraystretch}{0.5}
		\begin{tabular}{@{}*{8}{c@{\hspace{0.5mm}}}}
			{\bf \em h}&{\bf \em g}&{\bf \em b}&{\bf \em e}&{\bf \em a}&{\bf \em f}&{\bf \em c}&{\bf \em d}\\[0.5mm]

0&0&0\\
1&1&0\\
&0&0&1\\
&1&0&1\\
&&0&1&0\\
&&0&1&1\\
&&&1&0&1\\
&&&1&1&1\\
&&&&0&1&1\\
&&&&1&1&1\\
&&&&&1&1&0\\

		\end{tabular}}

};

	\node (nHFourTwo) [right=of nJFourTwo] {

		{\renewcommand{\arraystretch}{0.5}
		\begin{tabular}{@{}*{8}{c@{\hspace{0.5mm}}}}
			{\bf \em h}&{\bf \em g}&{\bf \em b}&{\bf \em e}&{\bf \em a}&{\bf \em f}&{\bf \em c}&{\bf \em d}\\[0.5mm]

0&0&0\\
1&1&0\\
&0&0&1\\
&1&0&1\\
&&0&1&0\\
&&0&1&1\\
&&&1&0&1\\
&&&1&1&1\\
&&&&0&1&1\\
&&&&1&1&1\\
&&&&&1&1&1\\

		\end{tabular}}

};

	\node (nHFourThree) [right=of nJFourThree] {

		{\renewcommand{\arraystretch}{0.5}
		\begin{tabular}{@{}*{8}{c@{\hspace{0.5mm}}}}
			{\bf \em h}&{\bf \em g}&{\bf \em b}&{\bf \em e}&{\bf \em a}&{\bf \em f}&{\bf \em c}&{\bf \em d}\\[0.5mm]

0&0&0\\
1&1&0\\
&0&0&1\\
&1&0&1\\
&&0&1&0\\
&&&1&0&0\\
&&&&0&0&1\\
&&&&&0&1&1\\

		\end{tabular}}

};

	\node (nHFiveOne) [left=0.8cm of nJFiveOne] {
		{\renewcommand{\arraystretch}{0.5}
		\begin{tabular}{@{}*{8}{c@{\hspace{0.5mm}}}}
			{\bf \em h}&{\bf \em g}&{\bf \em b}&{\bf \em e}&{\bf \em a}&{\bf \em f}&{\bf \em c}&{\bf \em d}\\[0.5mm]

0&0&0\\
&0&0&1\\
&&0&1&0\\
&&0&1&1\\
&&&1&0&1\\
&&&1&1&1\\
&&&&0&1&1\\
&&&&1&1&1\\
&&&&&1&1&0\\
&&&&&1&1&1\\

		\end{tabular}}
};

	\node (nHFiveTwo) [right=of nJFiveTwo] {
		{\renewcommand{\arraystretch}{0.5}
		\begin{tabular}{@{}*{8}{c@{\hspace{0.5mm}}}}
			{\bf \em h}&{\bf \em g}&{\bf \em b}&{\bf \em e}&{\bf \em a}&{\bf \em f}&{\bf \em c}&{\bf \em d}\\[0.5mm]

1&1&0\\
&1&0&1\\
&&0&1&0\\
&&&1&0&0\\
&&&&0&0&1\\
&&&&&0&1&1\\

		\end{tabular}}
};

}

\end{tikzpicture}
\end{center}
\caption{Hyperstructure}
\end{figure}
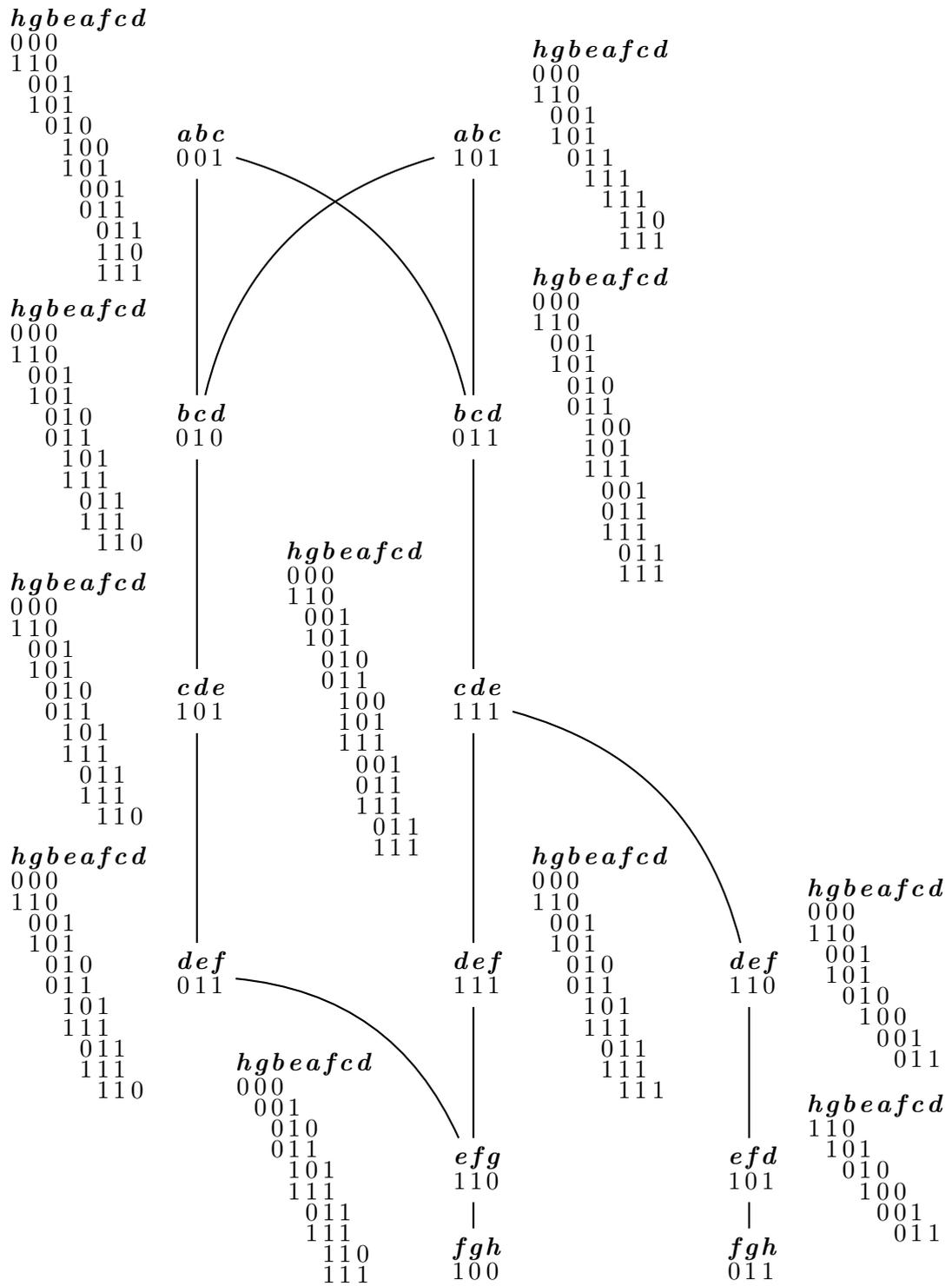

\newpage
Yet, the attainable requirement to substructure-vertex features may be formulated as
follows: each substructure-vertex at the $j$th tier must be a union of $j$-intersections and, consequently, each substructure-vertex at the $(n-2)$th tier must be a union of $(n-2)$-intersections. Then, evidently, the sufficient condition for the existence of JS sets will be satisfied. The following theorem terminates the given reasoning.

{\bf THEOREM 1}. JS sets for CT structures  $S_1$ and  $S_2$   exist if and only if the effective procedure forms a non-empty hyperstructure based on these structures.

{\bf PROOF}. The necessary condition has been proved by the STATEMENT.

The proof of the sufficient condition is based on the induction principle on $j$ that is the number of completed tiers. The statement formulation that is to prove concerns each vertex $v_j$ of the $j$th  tier of the HS, $j=1, 2, \ \dots,\ n-2$, and can be expressed as follows:
$$
\pi_j=\bigcup_{i=1}^{d_j} B_i^j\, ,\eqno (2)
$$
where $\pi_j$ is the substructure-vertex assigned to $v_j$; $B_i^j$ is the $i$th $j$-intersection of substructure-vertices; $d_j$ is  the number of  $j$-intersections that are united in $\pi_j$.

The wording of (2) is as follows: {\it the substructure-vertex $\pi_j$ is a union of  $j$-intersections}.

\underline{Induction basis}. It is evident that (2) is true for $j$ = 1, 2, 3 (note that each $\pi_j$, when $j\ge 3$, is calculated with the filtration procedure).

\underline{Induction step}. Suppose (2) is true for each $v_j$ (and, consequently, $\pi_j$) when $j\le t$. Then we show that (2) is true for each $v_j$ when $j=t+1$ $(3\le t \le n-3)$.

We will examine for the distinct vertex $v_{t+1}$ any adjacent vertex $v_t$. The calculation of the substructure-edge $\pi_{t,t+1}$   includes concretization of the variable $x_{t+3}$; let $\beta$ be the concretized value of $x_{t+3}$ ($\beta \in \{0,\ 1\}$). The substructure $\pi_t$, by the induction hypothesis, is the union of $t$-intersections $B_i^t$,\,  $i=1, 2, \ \dots,\ d_t$.

In $\pi_t$, two substructures can be distinguished and separated: $\pi_t(\beta)$ and $\pi_t(\overline{\beta})$ that unite, respectively, $t$-intersections with $x_{t+3}=\beta$ and $t$-intersections with  $x_{t+3}=\overline{\beta}$. It follows from the induction hypothesis that (2) is true for $v_t$ in a {\it special case} of identity $x_{t+3}\equiv \beta$ in all substructures that take part in forming  $\pi_t$. Hence, the separation of $\pi_t(\beta)$ can be obtained by the following means:

({\bf a}) concretization:  $\pi_t(x_{t+3}\to \beta)$;

({\bf b}) projection of the tiers 1, 2, \ \dots ,\  $t-1$ onto the substructure $\pi_t(x_{t+3}\to \beta)$.

In aggregate these points, taking into account REMARK 2 in this Section, realize the specified special case: calculation of  $t$-intersections with constant value $x_{t+3} = 0$.

But the operations fixed in ({\bf a}) and ({\bf b}) are exactly the same as the operations defined in EP for the substructure-edge $\pi_{t,t+1}$ calculation, consequently,  $\pi_{t,t+1}=\pi_t(\beta)$. On the other hand, according to definition of the substructure-edge,  $\pi_{t,t+1}=\pi_t\cap \pi_{t+1}$, consequently, the union of $(t+1)$-intersections in the form of $\pi_t(\beta)$ is shifted along the edge $(v_t,\,v_{t+1})$ and put in the place of the substructure-vertex  $\pi_{t+1}$.

The same argument applies to another edge $(v'_t,\,v_{t+1})$ that exists if the vertex $v_{t+1}$ has two adjacent vertices: $v_t$  and  $v'_t$. In this case $\pi_{t+1}$, again in accordance with definition, is the union of two substructure-edges and also the union of  $(t+1)$-intersections. Hence, (2) is true for all vertices in HS.

Thus, any non-empty substructure-vertex at the $(n-2)$th tier of HS is the union of $(n-2)$-intersections, and, what is the same, JS sets for $S_1$ and $S_2$. This conclusion terminates the proof of the sufficient condition for two CT structures. $\Box$

\medskip
{\bf 5. Solution of the general 3-SAT problem}

A general problem formulation that involves a modification of the effective procedure is as follows: given the $k$  unified CT structures $S_1$, $S_2$, \ \dots,\  $S_k$, it is required to ascertain the fact of existence (or absence) of JS sets.

We consider the graph $G_1$ associated with $S_1$ as a basic graph. The modified EP creates $k-1$ copies of $G_1$: $G_2$, \ \dots,\ $G_k$ . The system of CT structures and the enumerated graphs (as skeletons) are used for {\it parallel forming a system of hyperstructures} (HSS):  $\Gamma_2$ (on the basis of $S_1$ and $S_2$), $\Gamma_3$(on the basis of $S_1$ and $S_3$), \ \dots,\  $\Gamma_k$ (on the basis of $S_1$ and $S_k$). We denote this system as $\Gamma^\wedge$ = $\{\Gamma_2, \ \dots ,\ \Gamma_k\}$.

Each HS $\Gamma_r$, $r=2, \ \dots,\ k$, is formed according to the items formulated  in Section 4, with additional {\it concordance rules}. Let us postulate that the term ``same-name'' is applicable to all substructures (both intermediate and resulting in computation process) that are processed {\it in parallel} in
hyperstructures $\Gamma_2, \ \dots,\ \Gamma_k$, if these substructures are assigned to the same-name copies of some vertex or to the same-name copies of some edge of the basic graph. The substructures generated by the same-name substructure-vertices in the course of parallel computation (in particular, substructures that appear as intermediate results of filtration procedure) can also be classified as same-name objects.

We say that $j$-intersections in different hyperstructures are {\it same-name}, if the substructures forming each $j$-intersection are assigned to a distinct set of vertices in terms of the basic graph. The parallel processing of hyperstructures determines the strict operation order: every new step of the modified EP is executed only when the preceding step is completed for each HS $\Gamma_r$, $r=2, \ \dots,\ k$. The term {\it step} denotes creation or modification of any substructure (and, consequently, of all same-name substructures in the HSS).

The concordance rules are as follows.

\hangindent=1.1cm
A)\  All same-name substructures are to be unified in the course of the HSS parallel forming.

\hangindent=1.1cm
B)\  A {\it concordant shift} of the substructure-vertex $\pi_j$ of the  $j$th tier along the edge $(v_j,\,v_{j+1})$ in each HS: $\Gamma_2, \ \dots,\ \Gamma_k$, is based on the shift that is defined in Section 4 for a single hyperstructure with addition of unification as a crucial point.

\hangindent=1.1cm
C)\  In the course of the HSS forming, all BG vertices and edges that have no copies in the HS system are to be removed from BG.

Thus, the hyperstructures $\Gamma_2, \ \dots,\ \Gamma_k$ can be empty or not empty only jointly. The EP that realizes the concordance rules for $\Gamma^\wedge$ will be called a {\it systemic effective procedure} (SEP).

\medskip
{\bf THEOREM 2}. JS sets for CT structures $S_1,\,S_2, \,\ldots,\, S_k$ exist if and only if the SEP forms a non-empty system  $\Gamma^\wedge$ of hyperstructures based on these CT structures, hence, the formula $F$ is satisfiable if and only if $\Gamma^\wedge \ne \oslash$.

{\bf PROOF}. The necessary condition is evident since if $\Gamma^\wedge = \oslash$, then JS sets associated with the same-name $(n-2)$-intersections in $\Gamma_2, \ \dots,\ \Gamma_k$ cannot exist.

The proof of the sufficient condition is based on the induction principle on $j$ that is the number of completed tiers. The statement formulation that is to prove concerns any same-name vertices $v_j$
at the $j$th tier, $j=1, 2, \ \dots,\ n-2$, in hyperstructures $\Gamma_2, \ \dots,\ \Gamma_k$ and can be expressed as follows:

$$
\pi_j(\Gamma_r)=\bigcup_{i=1}^{d_j(\Gamma_r)} B_i^j(\Gamma_r),\;r=2, \,\ldots,\, k\,,\eqno (3)
$$

where:

$$
B_i^j(\Gamma_2)\leftrightarrow B_i^j(\Gamma_3)\leftrightarrow \cdots
\leftrightarrow B_i^j(\Gamma_k)\,,\eqno (4)
$$
$$
d_j(\Gamma_2)=d_j(\Gamma_3)= \cdots =d_j(\Gamma_k)\,.\eqno (5)
$$

The notation used in (3)--(5):

\medskip
$\pi_j(\Gamma_r)$ \hspace{0.5cm} same-name substructures assigned to $v_j$ in $\Gamma^\wedge$, $r=2, \ \dots,\ k$;

\smallskip
$B_i^j(\Gamma_r)$ \hspace{0.4cm} $i$th  $j$-intersection of substructure-vertices in  $\Gamma_r$;

\smallskip
$\leftrightarrow$ \hspace{1.2cm} symbol ``same-name'' for  $j$-intersections written on both sides of it;

\smallskip
$d_j(\Gamma_r)$ \hspace{0.5cm}	number of  $j$-intersections that are united  in  $\pi_j(\Gamma_r)$.

\smallskip
The wording of (3) is as follows: {\it the same-name substructure-vertices of $j$}th {\it tier in $\Gamma_2, \ \dots,\ \Gamma_k$   are unions of same-name $j$-intersections}.

\smallskip
\underline{Induction basis}. It is evident that (3)--(5) are true for  j = 1, 2, 3.

\smallskip
\underline{Induction step}. Suppose (3)--(5) are true for each $v_j$
(and, consequently, $\pi_j$) when $j\le t$. Then we show that (3)--(5) are true for each $v_j$ when $j=t+1$ $(3\le t \le n-3)$.

We will examine for the distinct $v_{t+1}$ in each $\Gamma_r$, $r=2, \,\ldots,\, k$, any adjacent vertex $v_t$ . The calculation of the substructure-edge $\pi_{t,t+1}$ includes concretization of the variable $x_{t+3}$; let $\beta$  be the concretized value of $x_{t+3}$  ($\beta \in \{0,\ 1\}$). The substructure $\pi_t(\Gamma_r)$, by the induction hypothesis, is the union of $t$-intersections $B_i^t(\Gamma_r)$, $i=1,\,2, \,\ldots,\, d_t(\Gamma_r)$. In $\pi_t(\Gamma_r)$ two substructures can be distinguished and separated:  $\pi_t(\Gamma_r, \beta)$ and $\pi_t(\Gamma_r, \overline{\beta})$ that unite, respectively, $t$-intersections with  $x_{t+3}=\beta$ and  $t$-intersections with $x_{t+3}=\overline{\beta}$. It follows from the induction hypothesis that (3)--(5) are true for  $v_t$  in a {\it special case} of identity   $x_{t+3}\equiv \beta$  for all substructures that take part in forming   $\pi_t(\Gamma_r)$. Hence, taking into account REMARK 2 in Section 4, the separation of  $\pi_t(\Gamma_r, \beta)$  can be obtained by the following means:

\hangindent=1.3cm
(a) concretization: $\pi_t(x_{t+3}\to \beta)$;

\hangindent=1.3cm
(b) projection of the tiers 1, 2,\, \ldots, $t-1$ onto the substructure  $\pi_t(x_{t+3}\to \beta)$, with unification of all intermediate and resulting substructures.

But in accordance with the rules of the shift and the concordant shift
fixed for SEP, $\pi_{t,t+1}(\Gamma_r)= \pi_t(\Gamma_r, \beta)$, hence (3)--(5) are true for $\pi_{t,t+1}(\Gamma_r)$, $r=2, \ \dots,\ k$. On the other hand, according to definition of the substructure-edge, $\pi_{t,t+1}(\Gamma_r)=\pi_t(\Gamma_r)\cap \pi_{t+1}(\Gamma_r)$, consequently, the union of the same-name $(t+1)$-intersections in the form of $\pi_t(\Gamma_r, \beta)$ is shifted along the edge $(v_t,\,v_{t+1})$ and put in the place of the substructure-vertex  $\pi_{t+1}(\Gamma_r)$.

The same argument applies to another edge marked as ($v'_t, v_t+1$) in hyperstructures $\Gamma_2, \ \dots,\ \Gamma_k$, that exists if the vertex $v_{t+1}$  has two adjacent vertices: $v_t$  and  $v'_t$. In this case the substructure $\pi_{t+1}(\Gamma_r)$,  $r=2, \ \dots,\ k$, is the union of two substructure-edges and also the union of  $(t+1)$-intersections. Thus, (3)--(5) are true for all vertices in the HSS.

So, any non-empty substructure-vertex at the $(n-2)$th tier is the union of $(n-2)$-intersections, coincident for same-name structures in the HS system, and, what is almost the same, the union of JS sets for $S_1,\, S_2, \,\ldots,\, S_k$. This conclusion terminates the proof of the sufficient condition for the general case. $\Box$

The brief essential interpretation of the presented proof is as follows: if  $F$  is a satisfiable formula, then in $\Gamma_2, \ \dots,\ \Gamma_k$ the equivalent sets of routes exist in terms of vertices and edges of BG. Hence, for each CTS  $S_r$,  $r=2, \ \dots,\ k$, there exists a subset of satisfying sets that is bijectively mapped onto the unique subset of satisfying sets for CTS  $S_1$.

The concrete JSS ascertaining the formula  $F$ satisfiability can be found by calculation of some $(n-2)$-intersection, performing a movement in HSS from any substructure of the $(n-2)$th tier back to the tier 1 with successive determination of the variables' values in the converse order.
\begin{table}[ht]
\begin{tabular}{|c|c|c|c|c|c|c|c|c|c|c|c|c|c|c|c|c|c|c|c|c|c|c|c|c|c|}
\multicolumn{26}{c}{\bf Table 5. Unified CT structures \,$S_1$, $S_2$,  $S_3$}\\
\multicolumn{26}{c}{$S_1$\hspace{5cm}$S_2$\hspace{5cm}$S_3$}\\
\cline{1-8}
\cline{10-17}
\cline{19-26}
$a$&$b$&$c$&$d$&$e$&$f$&$g$&$h$&&$h$&$g$&$b$&$e$&$a$&$f$&$c$&$d$&&
$d$&$f$&$a$&$c$&$h$&$e$&$b$&$g$\\
\hhline{|=|=|=|=|=|=|=|=|~|=|=|=|=|=|=|=|=|~|=|=|=|=|=|=|=|=|}
0&0&1&&&&&& &0&0&0&&&&&& &1&0&0&&&&&\\
\cline{1-8}
\cline{10-17}
\cline{19-26}
1&0&1&&&&&& &1&1&0&&&&&& &1&1&1&&&&&\\
\cline{1-8}
\cline{10-17}
\cline{19-26}
&0&1&1&&&&& &&0&0&1&&&&& &&0&0&1&&&&\\
\cline{1-8}
\cline{10-17}
\cline{19-26}
&&1&1&1&&&& &&1&0&1&&&&& &&1&1&1&&&&\\
\cline{1-8}
\cline{10-17}
\cline{19-26}
&&&1&1&0&&& &&&0&1&0&&&& &&&0&1&1&&&\\
\cline{1-8}
\cline{10-17}
\cline{19-26}
&&&1&1&1&&& &&&0&1&1&&&& &&&1&1&0&&&\\
\cline{1-8}
\cline{10-17}
\cline{19-26}
&&&&1&0&1&& &&&&1&0&0&&& &&&&1&1&1&&\\
\cline{1-8}
\cline{10-17}
\cline{19-26}
&&&&1&1&0&& &&&&1&1&1&&& &&&&1&0&1&&\\
\cline{1-8}
\cline{10-17}
\cline{19-26}
&&&&&0&1&1& &&&&&0&0&1&& &&&&&1&1&0&\\
\cline{1-8}
\cline{10-17}
\cline{19-26}
&&&&&1&0&0& &&&&&1&1&1&& &&&&&0&1&0&\\
\cline{1-8}
\cline{10-17}
\cline{19-26}
\multicolumn{8}{}{}    &&&&&&&0&1&1& &&&&&&1&0&1\\

\cline{10-17}
\cline{19-26}
\multicolumn{8}{}{}    &&&&&&&1&1&1& &&&&&&1&0&0\\
                                                        \cline{10-17}
\cline{19-26}
\end{tabular}
\end{table}

\medskip
The procedure of JSS determination for the given example uses the unified CT structures  $S_1$, $S_2$ and $S_3$ presented in Table 5 (Table 5 is based on the structures from Table 3). At forming the first tier of hyperstructures $\Gamma_2$ and $\Gamma_3$, the unification of substructures assigned to vertices marked as 001 and 101 (for  $a$, $b$, $c$  triplet) leads to obtaining two elementary CT structures corresponding to sequences 00111011 and 10111100 for initial variables numeration. In such cases a simple examination is necessary: whether the basic CTS $S_1$ contains at least one of the received sequences. The positive answer means satisfiability of the formula $F$. In the given example $S_1$ contains both sequences and, hence, they are JS sets. Thus the necessity for further HSS forming disappears.

\medskip
\textbf {6. Complexity of computation}

\smallskip
The maximal dimension of hyperstructures in HSS is a function of the number of variables in the initial formula $F$. Each hyperstructure contains no more than $8(n-2)$ substructures, each substructure includes no more than $8(n-2)$ lines. Concerning the operations on CT structures introduced in Section 2, we state that unification is the most complicated one: Its complexity is proportional $n^2$ because it involves searching through ${n\choose 2}$ combinations; the others are linear of  $n$ because they realize processing of the ordered lines of separate tiers.

Forming the tier numbered  $j$  in each HS is based on processing the substructures located at the tiers $1, 2, \ \dots,\ j-1$. This means that the real number of processed substructures is quadratic
dependent on $n$, taking into account the equation $1+2+ \cdots +(n-3) = (n-2)(n-3)/2$. Hence, the asymptotic estimation for the complexity of HSS forming is $O(n^4k)$. In terms of the problem parameters that determine an {\it input size} this expression should be transformed to $O(n^4m)$, because $k\le m$ (in most cases, $k\ll m$).
Thus, generally, the algorithm complexity is a polynomial function of the input size.

\medskip
\textbf{7. Algorithm testing and conclusions}

\smallskip
The computer-aided experiment consisted of two parts carried out at different periods of the research.

\smallskip
Part 1. Analysis of the program run-time as a function of input size.

The program realizing SEP was tested with the use of 2--3 GHz computers at parameter settings varied in intervals: $n=5\div 100$, $m=10\div 3000$.  The following modes of  data input and formula creation were used: keyboard input, file input, formula generation using a random number generator (RNG), with preassigned parameters ($n$, $m$, negation percentage for variables) and preassigned properties (optional, a priory satisfiable, a priory non-satisfiable). The main experimental
results presented more than 1000 testing runs for formulas with parameters: $n > 25$, $100 \le m \le 1000$, including  $n = 100$,  $m = 1000$ in a pair. The typical computing time values (in minutes) were: $\tau<1$ for $n\le 50$; $\tau=1\div 3$ for $n=64$; $\tau=5\div 8$ for $n=80$; $\tau=20\div 25$ for $n=100$.

The testing on the whole indicated a hundred percent successful formula classifications. The three possible messages for each individual formula were put in the program:

\hangindent=1.1cm 
1) {\it the formula is not satisfiable}---in case of the HSS forming failure, with an empty tier number indication (see the necessary condition of Theorem 2);

\hangindent=1.1cm
2) {\it the formula is satisfiable}---such a message was accompanied by obligatory presentation of the verified satisfying set;

\hangindent=1.1cm
3) {\it failure of classification}---in case when HSS forming was completed ($\Gamma^\wedge \ne \oslash$), but a satisfying set couldn't be found.

The first or the second messages terminated each run of the program, the third message didn't occur at all.

\smallskip
Part 2. Statistic testing of the algorithm adequacy and precision.

In this part of experiment 15 computers with MP clock rate exceeding 2 GHz were in use.
The program based on SEP was tested on the general totality of 410000 formulas; the parameter
intervals were: $n=20\div 45$, $m=100\div 300$. About 40\% of formulas were formed with
$n = 30\div 45$, with group average  $n_a \approx 38$. The regions of values for pair combinations ($n$, $m$) leading in most cases to a satisfiable or non-satisfiable formulas were determined in special series of experiments, in order to secure (approximately) an equal percentage of these two classes.

The results of this part of the experiment also showed a hundred percent successful classifications: each formula of 410000 was declared satisfiable or non-satisfiable (according to the first two messages), no failure of classification occurred.

In summary, the experimental results were in full accord with the theoretical foundation of the model presented in this paper.

The novel method used for 3-SAT problem resolution has been called a {\it bijective mapping principle for sets of components of discordant structures to a basic set}. The term {\it discordant} characterizes structures that do not come into operations with each other, except for unification.
The mapped components are $(n-2)$-intersections in a system of hyperstructures. The model as
a whole and the special constructive components are unique. By this reason, the paper doesn't refer to any preceding works of other authors, except for the fundamental works [1],  [2], and [3]; the reference to the previous version of the article in the electronic journal is also offered [4].

The results of the work assume a generalization by force of polynomial reducibility among intractable problems.

\begin{center}
{\bf ACKNOWLEDGMENT}
\end{center}

The author is grateful to Dr. D. I. Gusev and Dr. N. N. Zhebrun of VlSU, Vladimir, Russia, for their helpful comments and valuable suggestions on the earlier version of the paper and for verification of completeness of the algorithm description from the standpoint of its practical realization.

\begin{center}
{\bf REFERENCES}
\end{center}

[1]  Cook S. A. The Complexity of Theorem Proving Procedures, Proc. $3^{rd}$ ACM Symp. \hspace*{1.1cm} on the Theory of Computing,   
       ACM (1971), 151--158.

[2]  Garey M. R., Johnson D. S. Computers and Intractability: A Guide to the Theory of \hspace*{1.1cm}  NP-completeness. San Francisco:    
       W. H. Freeman \& Company, Publishers, 1979.

[3]   Karp R. M. On the Complexity of Combinatorial Problems, Networks, 5 (1975), 45--68.

[4]  Romanov V. F. Non-orthodox combinatorial models based on discordant structures~ //  \hspace*{1.1cm} Electronic Scientific Journal ``Investigated in Russia'', \\ \hspace*{1.1cm} \href{http://zhurnal.ape.relarn.ru/ articles/2007/143e.pdf}{http://zhurnal.ape.relarn.ru/articles/
2007/143e.pdf}

\end{document}